\pdfoutput=1 
\documentclass[8pt]{article}  \usepackage{times}
\usepackage{graphicx}

\usepackage{color}

\begin{document}

\twocolumn[{\LARGE \textbf{Influence of Lipid Heterogeneity and Phase Behavior on Phospholipase A$_2$ Action at the Single Molecule Level\\*[0.0cm]}}

{\large Martin Gudmand$^{1,2,\ast}$, Susana Rocha$^3$, Nikos S. Hatzakis$^4$,Kalina Peneva$^5$, Klaus M\"ullen$^5$, Dimitrios Stamou$^4$, Hiroshi Uji-I$^3$, Johan Hofkens$^3$, Thomas Bj\o rnholm$^{2,\ast}$ and Thomas Heimburg$^{1,\ast}$\\
{\small $^1$Membrane Biophysics Group, Niels Bohr Institute, University of Copenhagen, Denmark}\\
{\small $^2$Nano-Science Center (NSC), Department of Chemistry, University of Copenhagen, Denmark}\\
{\small $^3$Lab. Photochemistry and Spectroscopy, Dep. of Chemistry, KU Leuven, Belgium}\\
{\small $^4$Bio-Nanotechnology Laboratory, Department of Chemistry, University of Copenhagen, Denmark}\\
{\small $^5$Max Planck Institut f\"ur Polymerforschung, Ackermannweg 10, Mainz, Germany}\\

{\normalsize \textbf{ABSTRACT\hspace{0.5cm} We monitored the action of phospholipase A$_2$ (PLA$_2$) on L- and D-dipalmitoyl phosphatidylcholine (DPPC) Langmuir monolayers by mounting a Langmuir-trough on a wide-field fluorescence microscope with single molecule sensitivity. This made it possible to directly visualize the activity and diffusion behavior of single PLA$_2$ molecules in a heterogeneous lipid environment during active hydrolysis. The experiments showed that enzyme molecules adsorbed and interacted almost exclusively with the fluid region of the DPPC monolayers. Domains of gel state L-DPPC were degraded exclusively from the gel-fluid interface where the build-up of negatively charged hydrolysis products, fatty acid salts, led to changes in the mobility of PLA$_2$. The mobility of individual enzymes on the monolayers was characterized by single particle tracking (SPT). Diffusion coefficients of enzymes adsorbed to the fluid interface were between 3 $\mu$m$^2$/s on the L-DPPC and 4.6 $\mu$m$^2$/s on the D-DPPC monolayers. In regions enriched with hydrolysis products the diffusion dropped to $\approx$ 0.2 $\mu$m$^2$/s. In addition, slower normal and anomalous diffusion modes were seen at the L-DPPC gel domain boundaries where hydrolysis took place. The average residence times of the enzyme in the fluid regions of the monolayer and on the product domain were between $\approx$ 30 and 220 ms. At the gel domains it was below the experimental time resolution, i.e. enzymes were simply reflected from the gel domains back into solution.}\\*[0.0cm] }}
]
\setlength{\parindent}{0cm}
\footnotesize {$^{\ast}$Martin Gudmand made the experimental work. Corresponding authors: T. Bj\o rnholm, E-mail: tb@nano.ku.dk and T. Heimburg, E-mail: theimbu@nbi.dk. }\\*[0.1cm]
\footnotesize { \textbf{keywords:} single particle tracking, wide field microscopy, phospholipase A$_2$, domain formation, diffusion, monolayers}\\
\footnotesize { \textbf{Abbreviations:} PLA$_2$: phospholipase A$_2$; DPPC: 1,2-dipalmitoyl-sn-glycero-3-phosphocholine; PDI: perylene diimide; TRITC-DHPE: N-(tetramethyl{-}rho{-}damine-6-thiocarbamoyl)-1,2-dihexadecanoyl-sn-glycero-3-phosphoethanola{-}mine; R6G:  rhodamine 6G ; NHS ester: N-Hydroxisuccinimide ester; }\\

\normalsize

\section*{Introduction}
The interplay between enzymes and the cell membrane is essential for the regulation of a wide range of biological processes. Phospholipase A$_2$ (PLA$_2$) enzymes play an important role in these regulatory processes as they interact directly with the membrane by altering both its chemical composition and physical state, thereby controlling its function. By catalyzing the hydrolysis of sn-glycero-3-phospholipids at the sn-2 ester bond they release 1-lysophosphatidylcholine (lyso-PC) and a free fatty acid, e.g., arachidonic acid, which takes part in cell signaling \cite{Six2000}.\\

Secreted PLA$_2$ constitute small (14 kDa) calcium dependent proteins found ubiquitously in the extracellular space of mammalians; e.g. blood and tear fluid \cite{Saari2001}, as well as in insect venoms \cite{Six2000}. Previous studies have shown that the activity of both type IB (from, e.g., pancreas) and type IIA (from, e.g., snake venom) PLA$_2$Õs are highly dependent on the state and composition of the lipid system \cite{Grainger1989, Honger1996, Burack1997, Nielsen2002}. For instance, the activity towards phosphocholine (PC) vesicles is at a maximum near the gel-fluid phase transition temperature \cite{Honger1996, OpDenKamp1974}. At this temperature, gel and fluid state lipids coexist, and there is a maximum in both lipid state fluctuations and lipid lateral compressibility \cite{Honger1996, Heimburg2007a}.  The change in composition caused by lipid hydrolysis effects the phase behavior of the membranes. Since biological membranes exist in a state slightly above their melting transition, phospholipases seem to play an important role in maintaining this state  \cite{Heimburg2007a}.
A further intriguing feature of PLA$_2$ kinetics is the frequently observed lag-burst phenomenon. This is especially pronounced on zwitterionic substrates, such as PCÕs. It is characterized by slow initial hydrolysis (the lag period) followed by a sudden increase in activity by several orders of magnitude (the burst) \cite{Honger1996, ApitzCastro1982, Nielsen1999}. Various studies have shown that the lag phase can be practically annihilated by addition of hydrolysis products  \cite{Honger1996, ApitzCastro1982, Nielsen1999}. This led to the notion that the burst is induced by phase separation (domain formation) of products accumulated in the membrane during the lag period. Furthermore, it has been suggested that the presence of negatively charged hydrolysis products (e.g. ionized free fatty acids) increases the electrostatic binding between the positively charged surface patch on PLA$_2$ (referred to as the \textit{i-face}) in which the entrance to the active site is located \cite{Mouritsen2006, Leidy2006}. Based on such observations a Òsubstrate theoryÓ has evolved in which the thermodynamic state of the lipid structure, rather than the molecular structure of the individual lipids, determines the overall enzymatic activity \cite{Mouritsen2006}. \\

Numerous studies have shown that PLA$_2$ is interfacially activated meaning that it only shows activity towards aggregated lipid structures, while it is virtually inactive on lipid monomers in solution \cite{Verger1976}. Molecular dynamics modeling suggests that even when tight binding is assumed between the enzyme and the phospholipid substrate, i.e. with PLA$_2$ partially penetrating the lipid structure, a distance of $\approx$1.5 nm from the outer plane of the lipid structure to the active site of the enzyme still remains \cite{Zhou1996}. This indicates a need for considerable protrusion of the individual lipid molecules from the aggregated structure in order to fit into the active site of the enzyme. Both lipid protrusion and enzyme penetration must be expected to be favored when the lipid state is highly fluctuating and lateral compressibility is at a maximum.\\

Despite its recognized importance, very few studies have directly investigated the dependence of heterogeneities and domain formation in the lipid structure on the activation of PLA$_2$, e.g. \cite{Grainger1989, Leidy2006, Simonsen2008}. To the best of our knowledge, direct visualization of single PLA$_2$ molecules and quantification of the effect of the heterogeneous lipid environment has not been reported in the literature. It was the aim of this study to investigate, at the single molecule level, the adsorption, lateral diffusion, and lateral partitioning behavior of PLA$_2$ and correlate it directly to the microstructure of the phospholipid monolayers during hydrolysis. \\

In order to perform such a study, we developed a novel monolayer trough designed primarily to accommodate high numerical aperture (NA) microscope objectives (scheme on Fig 1, setup described in \cite{Gudmand2009}). In comparison to previously published fluorescence studies on doped monolayers, e.g. \cite{Grainger1989, McConlogue1997}, and labelled PLA$_2$ \cite{Dahmen1998}, the combination of the novel Langmuir trough and a wide field microscope with single molecule detection sensitivity used in this study resulted in a drastic increase in optical resolution, signal-to-noise ratio, and temporal resolution. As has been shown, single molecule experiments can reveal phenomena that are hidden from ensemble data and provide new insights into the influence of lipid heterogeneities on the action of lipolytic enzymes \cite{Rocha2009, Sonesson2007}. Phospholipid monolayers were chosen for this study for several reasons. Most importantly, phospholipids are the natural substrate for PLA$_2$, their monolayers at the air-water interface are well-characterized in the literature \cite{McConlogue1997, Maloney1993}, and they mimic the natural situation in which the enzymes operate. In addition, they are thermodynamically well-defined lipid systems in which several parameters (surface pressure $\Pi$, mean molecular area MMA, temperature T, etc.) can easily be controlled. Furthermore, they are readily visualized using standard fluorescence microscopy methods. \\

In this study, the action of PLA$_2$-IB on DPPC monolayers in the lipid state coexistence region was monitored using two complementary types of fluorescence labeling schemes. In a first set of experiments, L-DPPC monolayers were fluorescently labelled with a lipid fluorophore which partitions exclusively in the fluid regions. In a second set of experiments, only PLA$_2$-IB was marked with a small organic dye, a water soluble perylene diimide (PDI) \cite{Margineanu2004}, which has no influence on enzyme activity \cite{Peneva2008}. When using low concentrations, single enzymes molecules were tracked while diffusing on the monolayer. Both L-DPPC and D-DPPC monolayers were studied. Although the binding affinity of PLA$_2$ for both lipids is similar \cite{Pattus1979a}, the enzyme shows no activity for the D-DPPC layer, which therefore served as a non-hydrolyzable surface.


\section*{Materials and Methods}
\textbf{Chemicals:} L-DPPC and D-DPPC \textbf{(left and right handed enantiomers) }were from Avanti Polar Lipids (Cat. \#: 850355) and from Sigma-Aldrich (Cat. \#: 42566), respectively. Fluorescence label TRITC-DHPE was supplied by Invitrogen (Cat. \#: T1391). Purified porcine pancreas PLA$_2$ (Type IB) was provided by Novozymes A/S (Bagsv\ae rd, Denmark). All solvents were spectroscopic grade from Merck Chemicals. MilliQ-water ($>$ 18.0 M½/cm at 25 $^{\circ}$C) was purified on a desktop Millipore system and it was used for all steps involving water. Ultra pure salts for buffer (TRIS pH 8.9, 150 mM NaCl, 5mM CaCl$_2$) were purchased from Merck Chemicals. All chemicals were used as received. 

\textbf{Enzyme labeling:} The \textbf{N-Hydroxisuccinimide (NHS)}-ester of PDI (PDI-NHS) was synthesized and purified as previously described \cite{Peneva2008}. Conjugation of PDI-NHS to PLA$_2$ was performed using standard procedures for protein labeling: PDI-NHS was added in 25-fold excess to an enzyme solution in carbonate buffer (pH 8.0). The solution was then incubated at 4$^{\circ}$C for two hours to produce the dye-labelled enzyme (PLA$_2$-PDI). The PDI-NHS bound to the enzyme through the amino groups of the lysine residues. Removal of unreacted dye from the enzyme solution was accomplished by several ($\approx$20) size exclusion spin filtrations with 5 kDa filters. Successive spin filtrations were performed until the filtrate was free of unreacted dye. The activity of the enzyme was verified on our monolayer setup with no detectable loss of activity. Absorbance spectroscopy showed on average three labels per protein. Fluorescence correlation spectroscopy (FCS) measurements confirmed the presence of a single diffusing species (PLA$_2$-PDI) with D$_{3D}$ = 100 $\mu$m$^2$/s (measured relative to R6G with known D$_{3D}$ = 300 $\mu$m$^2$/s \cite{Webb1974}). 

\textbf{Langmuir film preparation:} A home-build Langmuir Teflon\textregistered \ trough with internal dimensions 150x50 mm equipped with two moveable barriers made from Delrin\textregistered \ was used for the experiments. The trough was designed for our inverted fluorescence microscopy setup. To accommodate the short working distance of our microscope objective (see below), the glass observation window in the centre of the trough was elevated $\approx$3 mm above the Teflon trough bottom. This made it possible to keep a stable 3-4 mm high sub-phase in the majority of the trough, while minimizing the height of the sub-phase directly above the observation window of the required $\approx$200 $\mu$m. The Langmuir trough electronics and control software was from Kibron Inc. (Finland). A full description of the monolayer trough will be published elsewhere. To reduce surface flow of the monolayer in the observation region, a Teflon ring (\O \ 15 mm, height 3 mm) with a slit opening (2 mm) facing one of the barriers was placed on the cover-glass in the trough during experiments. A second cover glass was placed on top of the Teflon ring to further reduce air flow. The entire trough was covered with an acrylic case. Monolayers were spread from $\approx$0.6 mg/mL solution of lipid dissolved in hexane/ethanol 95/5 (v/v). Doped monolayers contained $<$ 0.1 mol\% TRITC-DHPE.
\begin{figure}[htb!]
    \begin{center}
	\includegraphics[width=8.5cm]{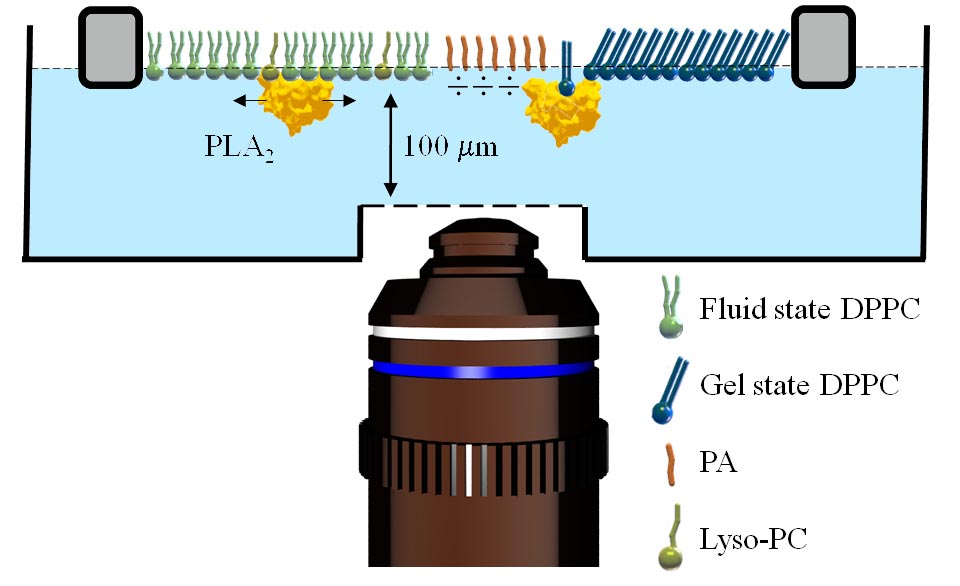}
	\parbox[c]{8cm}{ \caption{\textit{Schematic of the monolayer setup. A DPPC monolayer is compressed into the centre of the phase transition region ($\Pi$=8 mN/m, MMA = 65 \AA$^2$). At this surface pressure liquid state lipid molecules (green) coexist with gel state lipid molecules (blue), which form domains. In this coexistence region L-DPPC monolayers are susceptible to hydrolysis by PLA$_2$-IB (yellow). Hydrolysis leads to formation and accumulation of ionized free palmitic acid (PA, red) and lysophosphatidylcholine (lyso-PC, olive green) in the monolayer. The monolayer trough was custom-designed to accommodate the ~200 $\mu$m working distance of the high numerical aperture objective (NA = 1.2) and mounted on a home-build epi-fluorescence microscope. }
	\label{Figure1}}}
    \end{center}
\end{figure}

\begin{figure}[htb!]
    \begin{center}
	\includegraphics[width=8.5cm]{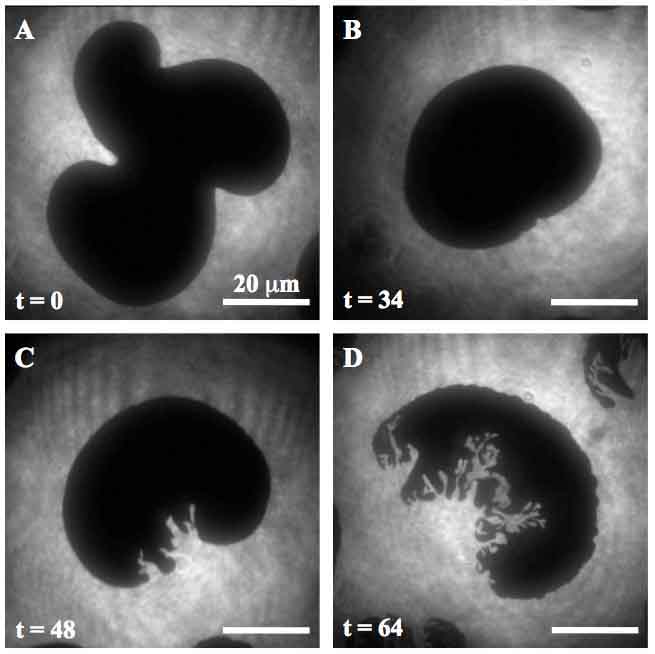}
	\parbox[c]{8cm}{ \caption{\textit{Time evolution of the morphology of a gel domain during PLA$_2$-IB catalyzed hydrolysis at time 0 min (A), 34 min (B), 48 min (C), and 64 min (D) after compression to 65 \AA$^2$. Note that the domain in image A is not the same as the domain followed in images B-D due to a slight drift in the monolayer. The L-DPPC monolayer was doped with the fluorescent lipid analogue (TRITC-DHPE). }
	\label{Figure2}}}
    \end{center}
\end{figure}

\textbf{Addition of enzyme beneath the monolayers:} Two different strategies were used for injection of enzyme. In experiments with TRITC-DHPE doped monolayers (Fig. \ref{Figure2}), the enzyme was dissolved in the sub-phase prior to spreading the monolayer and compression of the monolayer. This ensured a homogenous concentration in the aqueous sub-phase when the experiment was initiated. For experiments with fluorescently marked PLA$_2$ (PLA$_2$-PDI) the enzyme was injected, with a Hamilton syringe immersed from behind the monolayer barriers after monolayer compression. The enzyme was injected in proximity to the observation window. This created a gradient in the enzyme concentration near the observation area which made it possible to select areas with suitable surface density of enzyme, i.e. well-separated particles, as required for single particle tracking.

\textbf{Wide field microscopy:} Images presented in Fig. \ref{Figure2} were recorded on a CCD camera (Apogee KX85, pixel array: 1300 x 1030, pixel size: 6.7 $\mu$m). All other images, including image series for SPT, were recorded on an EMCCD camera from Andor (IXON EM$+$, DU897BV), pixel array: 512 x 512, pixel size: 16 $\mu$m). All images shown here were recorded using an Olympus 60x, water immersion, NA 1.2, UPLAPO objective (working distance: 0.22 mm). The Apogee camera was used in combination with a 1x camera lens. The Andor camera was used with a 2.5x camera lens (Total magnification: 150x). Samples were excited at 532 nm at a final excitation irradiance at the sample plane of 0.5 - 2 kW/cm$^2$. 
\begin{figure*}[htb!]
    \begin{center}
	\includegraphics[width=12cm]{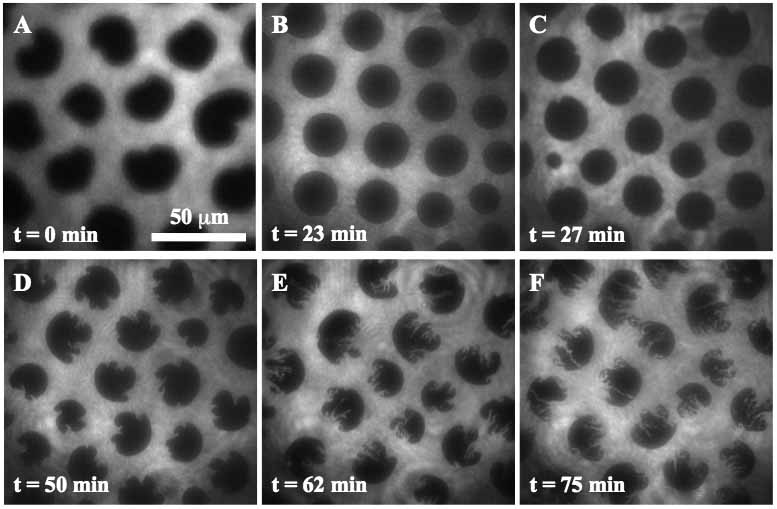}
	\parbox[c]{16cm}{ \caption{\textit{Time evolution of the morphology of several gel domains during PLA$_2$-IB catalyzed hydrolysis. (A) Domain shape after compression to $\approx$65 \AA $^2$. (B) Domains relaxed into the characteristic circular shapes observed before channel formation. (C) Onset of channel formation occurs simultaneously on all domains (D-F). The overall growth pattern of hydrolysis channels penetrating the domains appear identical on all domains. Note that due to a slight drift of the monolayer, the domains in the images A-F are not the same. The L-DPPC monolayer was doped with the fluorescent lipid analogue (TRITC-DHPE).}
	\label{FigureS1}}}
    \end{center}
\end{figure*}

\textbf{Single Particle tracking: }The determination of single enzyme trajectories was performed using a home-developed routine in Matlab\textregistered. The enzyme can be located with a precision of $\approx$100 nm for slowly diffusing enzymes and $\approx$200 nm for quickly diffusing enzymes. The values of the diffusion coefficient for all the different enzyme motions were determinate using cumulative distribution functions \cite{Rocha2009,Schutz1997}. 


\section*{Results and Discussion}
\textbf{Visualizing enzymatic action on fluorescently labelled monolayers:} In this series of experiments the monolayers were labelled with TRITC-DHPE. Using this fluorophore the gel L-DPPC domains appeared as dark regions, while fluid regions (L-DPPC + TRITC-DHPE) appeared bright in the fluorescence images. This allowed us to monitor domain structural changes during enzymatic activity. \\
\begin{figure*}[htb!]
    \begin{center}
	\includegraphics[width=12cm]{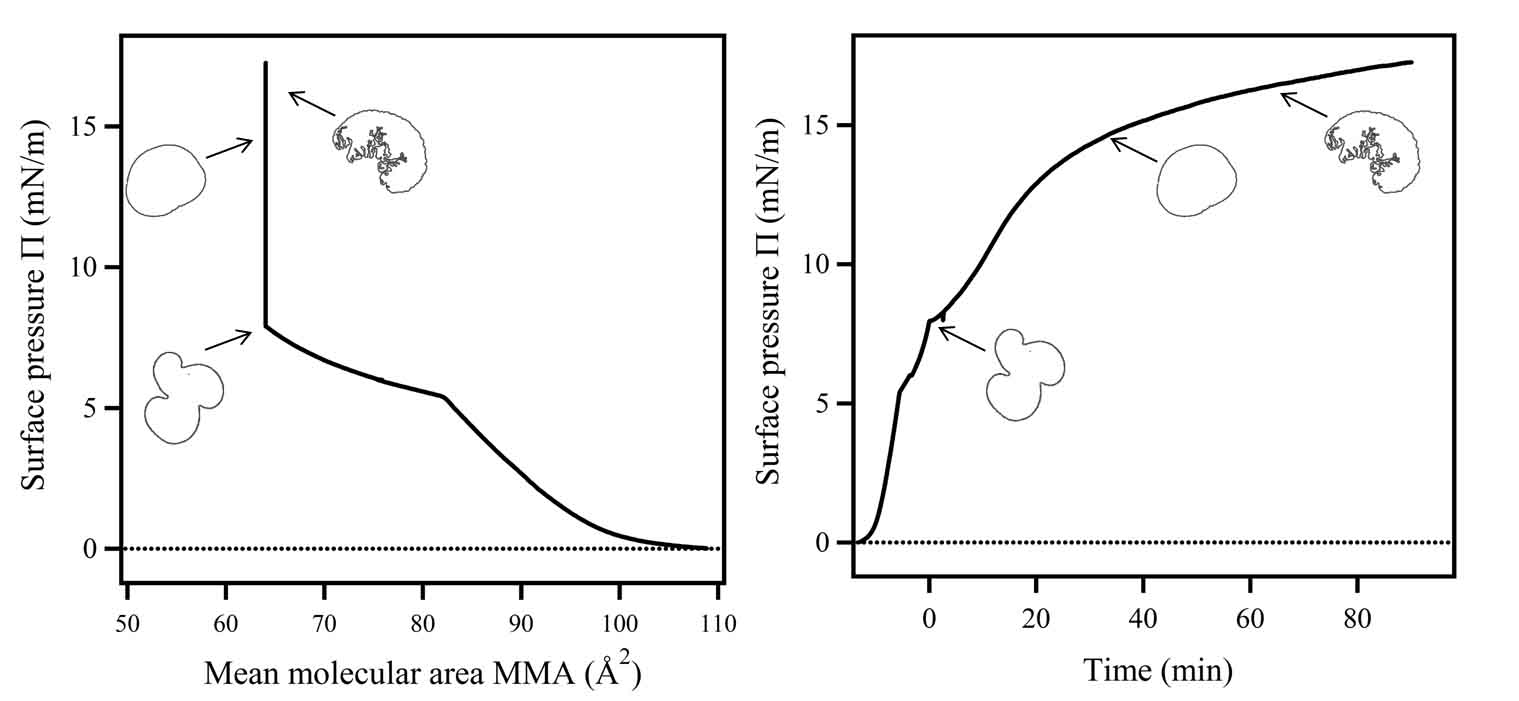}
	\parbox[c]{16cm}{ \caption{\textit{Left: Pressure-area isotherm from the experiment shown in Fig. 2. Compression was started at a MMA 110 \AA $^2$, and the onset of the phase transition region is seen at MMA 82-83 \AA $^2$. Compression was stopped at a target MMA of 65 \AA $^2$ corresponding to a surface pressure of 8 mN/m. The monolayer was kept a constant area during the enzyme adsorption and hydrolysis process. Domain shapes at corresponding times and pressures during the hydrolysis process are shown for reference. Right: Pressure-time plot: Compression was started at t= -13 min, the phase coexistence region was reached at t = -6 min, and compression was stopped at MMA 65 \AA $^2$ where large tri-lobed domains were formed (t$\equiv$0). }
	\label{FigureS2}}}
    \end{center}
\end{figure*}
The first indication of enzymatic action appeared as an accelerated relaxation of the bean shaped or tri-lobed L-DPPC domains into circular domains within a time-span of 30 minutes (Fig. \ref{Figure2}A-B). Relaxations from multi-lobed shapes are also seen in absence of enzyme, but in that case typically take several hours or days \cite{McConlogue1997, Klopfer1996}. The circular gel domains were then degraded from the gel-fluid interface as channels of fluid regions began to spread into the gel domains (Fig. \ref{Figure2}C). Strikingly the channel formation occurred simultaneously on practically all domains within a given experiment (see Fig. \ref{FigureS1}). This was always observed, even though the lag time before channel formation varied considerably in-between separate but identical experiments (tlag = 40±10 min). This is a strong indication of a thermodynamic control of the process since simultaneously occurring domain degradation must be controlled by macroscopic properties, and not by the microscopic structure of the individual domains. The overall growth pattern of the channels appeared similar on all domains (Fig. \ref{FigureS1}). Degradation furthermore occurred from only one side of the domains; leaving the opposing side remarkably inert to hydrolysis (Fig. \ref{Figure2}C-D). The channel growth appeared directed, and in many cases followed a straight line over large distances on the molecular level. Taken together this indicates that channel formation is linked to the internal structure of the L-DPPC gel domain. It seems likely that individual lipids in the gel domains are oriented on a lattice and can only be attacked by the enzyme from one side. The structure of the partly degraded domains, observed here for pancreatic PLA-IB, closely resemble those reported for snake venom PLA$_2$-IIA on a similar system \cite{Grainger1989, Grainger1990}. \\

The experiments were performed under constant area conditions to avoid disturbance introduced by the movement of the barriers. As a consequence, the lateral pressure, $\Pi$, typically increased from initially 8 mN/m to finally 17 mN/m (t = 91 min). Most likely, this was caused by both enzyme adsorption, enzyme penetration into the monolayer, and hydrolysis. Complete pressure-area and pressure-time isotherms associated with Fig. \ref{Figure2} are given in Fig. \ref{FigureS2}.\\
\begin{figure}[htb!]
    \begin{center}
	\includegraphics[width=8.5cm]{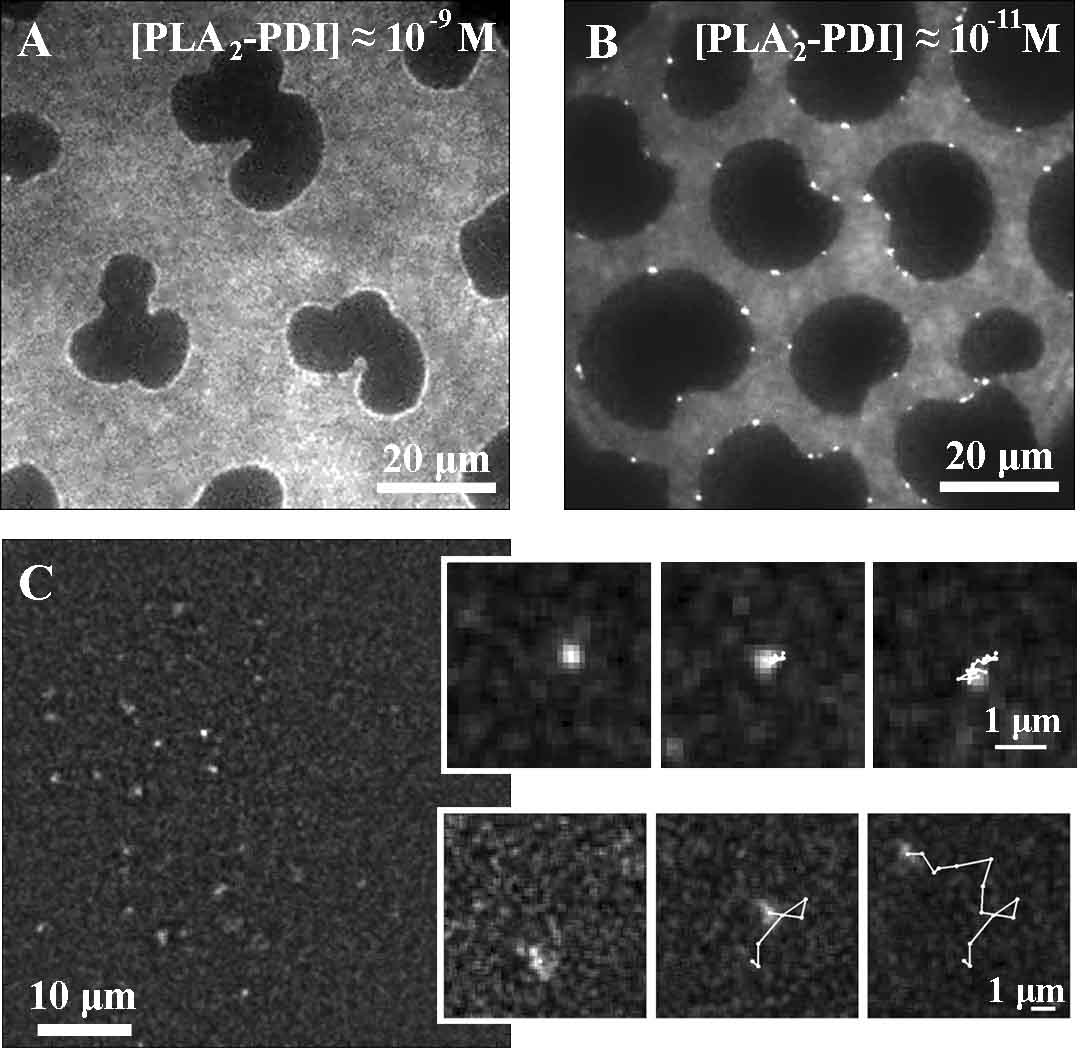}
	\parbox[c]{8cm}{ \caption{\textit{Wide-field fluorescence images of L-DPPC monolayers compressed to $\Pi$=8 mN/m (22 ¡C) with different concentrations of fluorescently labelled PLA$_2$ enzymes (PLA$_2$-PDI). (A-B) When the enzyme is present in high concentration, it is possible to visualize the domain structure (see text). It is also evident, from the bright regions along the domain interface, that a higher local concentration of PLA$_2$-PDI is found along the fluid/gel boundary. (C) At low concentration, it is possible to discriminate and track single enzyme molecules (fluorescence image after linear deconvolution process). The magnifications show trajectories described by slow diffusing (top) and fast diffusing (bottom) PLA$_2$-PDI molecules. Integration time: 22 ms.}
	\label{Figure3}}}
    \end{center}
\end{figure}

\textbf{Visualizing individual PLA$_2$-PDI enzymes: }In this set of experiments PLA$_2$-IB was labelled with the highly photo-stable organic dye PDI \cite{Margineanu2004, Peneva2008}. In this way, the location and mobility of labelled enzyme (PLA$_2$-PDI) could be tracked, analyzed, and linked to different lipid regions in the monolayer. Fig. \ref{Figure3} shows typical images of an undoped L-DPPC monolayer with different amounts of PLA$_2$-PDI added to the aqueous sub-phase (cf. Fig. \ref{Figure1}). All fluorescence contrast stems from labeled protein associated to the monolayer.\\

At relatively high concentrations of enzyme ($>$ 10$^{-11}$ M) the fluorescence images reveal the domain-segregated structure of the monolayer (Fig. \ref{Figure3}A and B) due to the different affinities of the enzyme towards regions of different lipid packing (i.e. lipid states). These images directly show that PLA$_2$-PDI interacts preferential with the fluid regions as evident from the bright interface of the domains. In principle, the image contrast should also make it possible to estimate the partition coefficient of PLA$_2$ between the fluid regions and the gel domains. In practice though, only a very low signal from the enzyme could be observed in the gel regions and therefore the enzyme is best characterized as having no affinity towards these dense gel state lipid domains. The bright regions observed along the domain interface are caused by a higher local PLA$_2$-PDI concentration and/or slower diffusion of the enzyme molecules at the fluid/gel boundary. At the highest concentration used (Fig. \ref{Figure3}A, [PLA$_2$-PDI] $\approx$ 10$^{-9}$ M), the entire fluid-gel interface is covered with PLA$_2$-PDI, albeit differences in fluorescence intensity witnessed at the liquid/gel interface indicate a non-uniform distribution of labelled PLA$_2$-PDI molecules. Thus, the hydrolysis of the domains from only one side of the domains (Figs. \ref{Figure2} and \ref{FigureS2}) cannot be due to different affinities of the enzyme at different sides of the domains. At lower concentration (Fig. \ref{Figure3}B, [PLA$_2$-PDI] $\approx$ 10$^{-11}$ M), the enzyme is clearly accumulated at discrete sites on the domain interface (bright spots), where it seems to be virtually immobilized. From the spot size and fluorescence intensity, we speculate that the majority of the spots are aggregates of several enzyme molecules. \\

In order to visualize the mobility of enzymes while acting on the phospholipid monolayers, the enzyme concentration was decreased to the picomolar regime. This was needed to ensure that the individual enzymes were well separated so that individual enzyme trajectories could be resolved. For this purpose, image time-series of monolayer regions with low surface density of the enzyme were recorded and analyzed (Fig. \ref{Figure3}C). As a result of the low concentration of enzyme, the monolayer structure could not be inferred from the individual image frames. \\

\textbf{Single particle tracking of PLA$_2$-PDI:} To investigate the influence of hydrolysis on the diffusion behavior of PLA$_2$ parallel experiments on L-DPPC and D-DPPC were performed at single molecule level. As pointed out previously, PLA$_2$ cannot hydrolyze D-DPPC and is thereby ÔinactiveÕ on the D-DPPC substrate monolayers. Nevertheless, PLA$_2$ has the same initial binding affinity for lipid structures of these two enantiomers \cite{Dahmen1998, Li2000b, Bonsen1972}. Although PLA$_2$ could be inactivated by removal of its co-factor Ca$^{2+}$ from the buffer, that approach was considered undesirable as Ca$^{2+}$ removal might affect the binding affinity of PLA$_2$ to the monolayer \cite{Pattus1979b} and would definitely have influenced the phase behavior and lipid packing of the monolayer structure itself \cite{Maloney1993}.\\
\begin{figure}[htb!]
    \begin{center}
	\includegraphics[width=8.5cm]{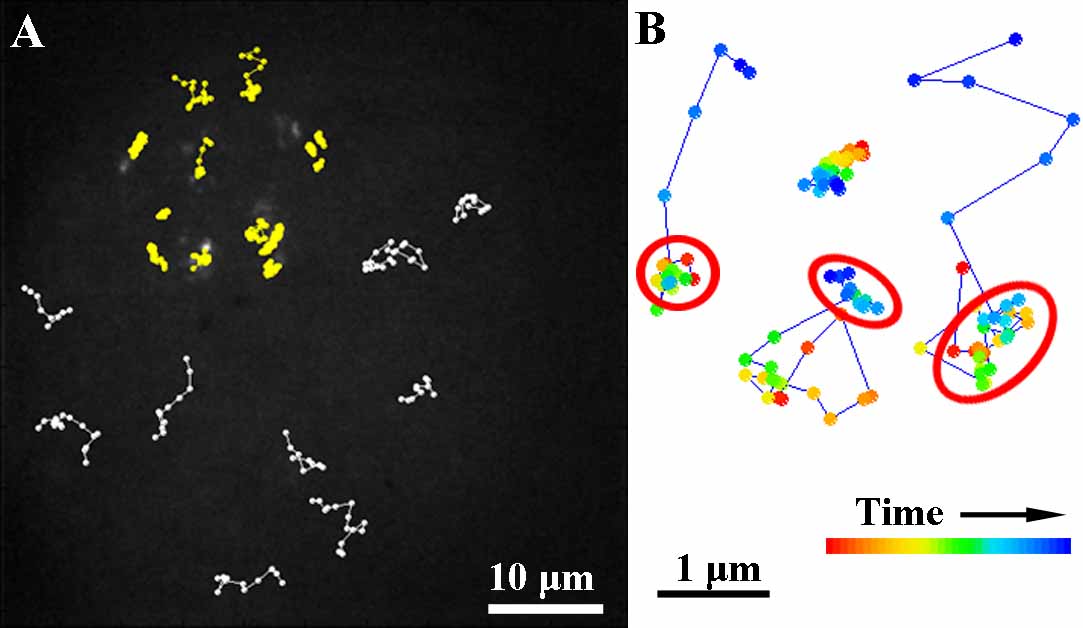}
	\parbox[c]{8cm}{ \caption{\textit{Diffusion behavior of labelled individual PLA$_2$ molecules. (A) Typical trajectories of individual PLA$_2$ molecules diffusing on the fluid region (white) and near the fluid/gel boundary (yellow). Background image accumulated over 100 frames. Four of the trajectories colored in yellow are magnified in (B). In three of the trajectories, it is possible to distinguish the hot spots (indicated by the red circles) where diffusion is slow.  
}
	\label{Figure4}}}
    \end{center}
\end{figure}
The advantage, and justification, of SPT experiments and analysis is its ability to detect heterogeneities in the trajectories followed by individual molecules. Single enzyme trajectories show that different PLA$_2$-PDI molecules may present different diffusion coefficients (Fig. \ref{Figure3}C). Moreover, when representative trajectories of single enzymes diffusing on L-DPPC monolayers are laid out, it is possible to discriminate two distinct spatial regions where phospholipase molecules diffuse differently (Fig. \ref{Figure4}A). While freely diffusing on the largest part of the monolayer in the field of view, on some regions the molecules seem to be immobilized or confined (e.g. top left corner of the image on Fig. \ref{Figure4}A). Taking into account the morphology that could be outlined at higher enzyme concentrations; we expect these immobilized or slowly diffusing PLA$_2$-PDI molecules to be located at the gel domain boundaries. \\

A detailed analysis of the trajectories followed by single enzyme molecules located near the fluid/gel boundary in Fig. \ref{Figure4} reveals heterogeneities both between and within the trajectories (Fig. \ref{Figure4}B). Four representative examples of enzymes that showed switching between fast and slow diffusion are depicted in Fig. \ref{Figure4}B. From the results obtained with the fluorescently labelled monolayer, we speculate that the trajectories detected in the vicinity of the gel domain region correspond to PLA$_2$-PDI molecules confined to the channels produced by hydrolysis (cf. Fig. \ref{Figure2}C and \ref{Figure2}D). Such transiently confined enzymes show switching between fast and slow diffusion and vice versa. The periods of slow diffusion are indicated by the red circles in Fig. \ref{Figure4}B. Importantly, this switching diffusion behavior was not detected for enzymes diffusing on the un-hydrolyzable D-DPPC monolayer.\\
\begin{figure}[htb!]
    \begin{center}
	\includegraphics[width=8.5cm]{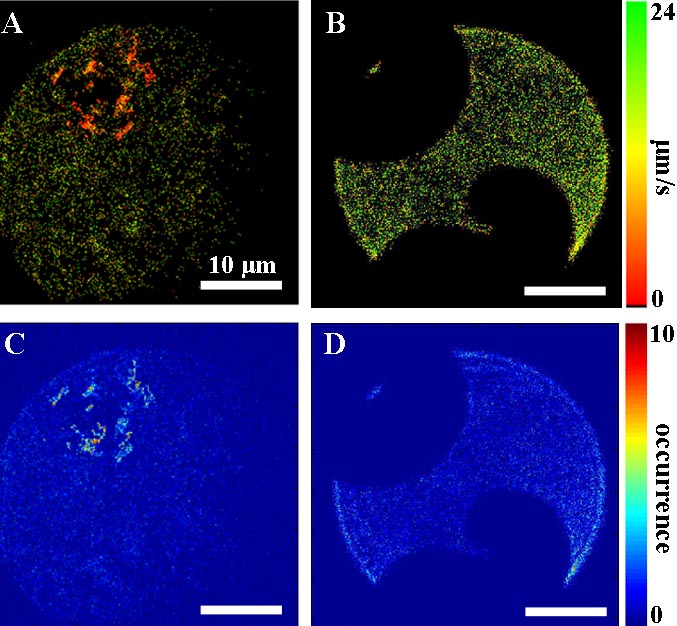}
	\parbox[c]{8cm}{ \caption{\textit{(A) M-LM image construct the mobility of PLA$_2$-PDI enzymes diffusing on a L-DPPC which is susceptible to hydrolysis. The enzyme diffuses markedly slower in distinct regions (red areas). (B) In the liquid regions of the non-hydrolyzable D-DPPC monolayer, measured enzyme diffusion coefficients are homogeneously distributed. (C-D) H-LM images showing the localization of PLA$_2$-PDI enzymes diffusing on a (C) L-DPPC and (D) a D-DPPC monolayer. Colored scale bars indicate measured diffusion coefficients and local occurrence of the enzyme. All scale bars in the images are 10 $\mu$m.}
	\label{Figure5}}}
    \end{center}
\end{figure}

The different mobility of the enzyme molecules on the different areas of the L-DPPC monolayer becomes clear when using Mobility-Localization Microscopy (M-LM) images. In all types of localization microscopy, e.g., PALM, STORM, PALMIRA, S-PALM \cite{Flors2007}, the position of single molecules within a frame is determined with high accuracy (depending on signal to noise, accuracies of up to 10 nm can be reached). This is done for all frames in the recorded movie and in the end a super resolution image is reconstructed. Here, instead of plotting the position of each molecule in the recorded movie on a final image, we plot the measured displacement of each individual enzyme between two consecutive image frames onto one final image. The images obtained this way map out the relation between the monolayer structure and the mobility of the enzyme. The two regions where PLA$_2$-PDI molecules present different diffusion behavior can now be identified clearly on the L-DPPC monolayer (Fig. \ref{Figure5}A), whereas enzyme motion on the fluid region of the D-DPPC monolayer is nearly homogeneous (Fig. \ref{Figure5}B). \\
\begin{figure}[htb!]
    \begin{center}
	\includegraphics[width=8.5cm]{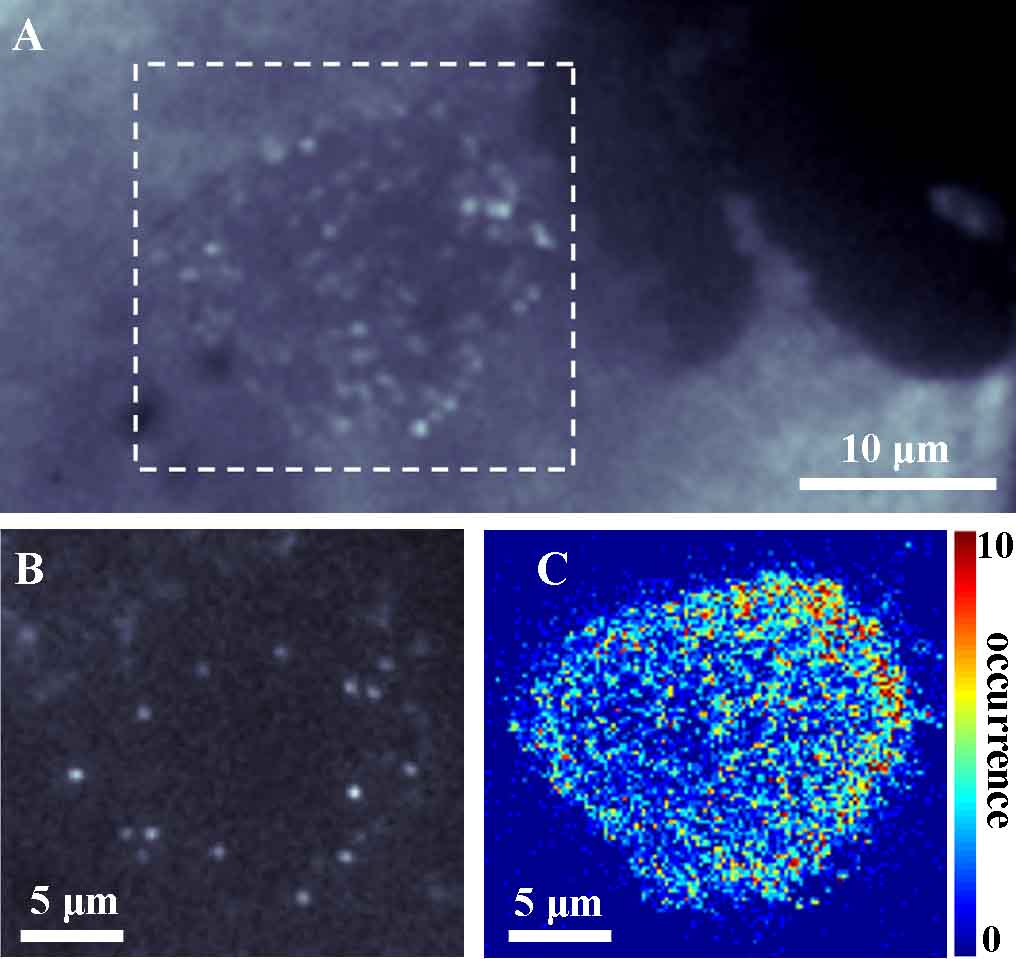}
	\parbox[c]{8cm}{ \caption{\textit{(A) Accumulation over 50 frames of a L-DPPC layer incubated with low enzyme concentration ($\approx$10-12M) for more than 60 minutes. It is possible to discriminate the gel domain (no enzymes, black region in upper right corner), the fluid region (enzymes diffusing fast leading to a uniform fluorescence) and the product domain (enzymes diffusing slowly or immobilized). The dashed square indicates the product domain region shown in B and C. (B) Fluorescence image of the product domain in which single enzyme molecules can be discriminated. Integration time: 30 ms. (C) The H-LM image shows a tendency for the enzyme to preferentially localize near the gel-fluid boundary.}
	\label{Figure6}}}
    \end{center}
\end{figure}
In a similar way the affinity of the enzyme for different regions of the monolayers can be evaluated by constructing Histogram-LM images (H-LM, \cite{Rocha2009}). H-LM images are constructed by re-plotting the detected localization of each individual enzyme from each image frame of a time series of images onto one final histogram image. The H-LM images for the PLA$_2$-PDI diffusing on the different monolayers are shown in Fig. \ref{Figure5}C and \ref{Figure5}D. \\

Although not immediately recognizable from the fluorescence images at low enzyme concentration, the gel domain structure are rendered visible both in M-LM and H-LM images (Fig. \ref{Figure5}). Especially on the D-DPPC monolayer where the domains are well defined and remain unchanged during time since no hydrolysis can occur. In the vicinity of the domains clear differences between the two systems are evident. On the L-DPPC monolayer the enzyme binds preferentially to localized spots, Ôhot-spotsÕ (shown in yellow and red) located at the gel-fluid boundary region (Fig. \ref{Figure5}A). In contrast, the enzyme is relatively homogeneously dispersed in the fluid region of the D-DPPC layer (Fig. \ref{Figure5}B). Interestingly, the Ôhot-spotsÕ coincide with the areas where enzyme diffuses slowly and they are limited to L-DPPC monolayers. We can therefore conclude that they are intimately linked to hydrolysis. This indicates that the regions of slow enzyme diffusion on L-DPPC are most likely hydrolysis product domains.\\
\begin{figure}[htb!]
    \begin{center}
	\includegraphics[width=8.5cm]{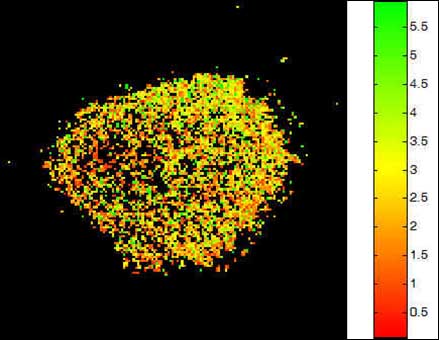}
	\parbox[c]{8cm}{ \caption{\textit{M-LM image of the product enriched area on L-DPPC (See also Fig. \ref{Figure6}). The image shows that the diffusion coefficient did not vary systematically within the product enriched area.
}
	\label{FigureS3}}}
    \end{center}
\end{figure}
The values of the diffusion coefficient for the different enzyme motions were determined using cumulative distribution functions \cite{Rocha2009, Schutz1997}. Within the fluid regions of the L-DPPC monolayer, all the enzymes seemed to diffuse randomly, with slightly different diffusion coefficients. In detail, the trajectories described by the majority of the molecules (87\%, 1216 molecules analyzed) exhibited only fast diffusion with D = 3.0 $\mu$m$^2$/s, while 13\% of the trajectories were found to contain both fast (D = 3.0 $\mu$m$^2$/s) and slow periods (D $<$ 0.26 $\mu$m$^2$/s). Conversely, near the gel-fluid boundary 16\% of the trajectories (515 molecules analyzed) exhibited slow diffusion (D = 0.025 $\mu$m$^2$/s). The majority of trajectories (61\%) showed a combination of slow (D = 0.025 $\mu$m$^2$/s) and fast (D = 0.27 $\mu$m$^2$/s) diffusion steps. The remaining 13\% of the molecules showed anomalous diffusion (D = 0.27 $\mu$m$^2$/s, $\alpha$ = 0.2) indicating confinement of the diffusion, probably in channels that are formed in the gel domain (see features in Fig. \ref{Figure2}). This behavior is in great contrast to the diffusion behavior found on D-DPPC. In the fluid region on D-DPPC 90\% of the trajectories (2971 molecules analyzed) exhibited normal fast diffusion (D = 4.6 $\mu$m$^2$/s). The remaining 10\% was heterogeneous having periods of slow (D $<$ 0.16 $\mu$m$^2$/s) and fast diffusion. Since diffusion is faster on D-DPPC (D = 4.6 $\mu$m$^2$/s) than on L-DPPC (D = 3.0 $\mu$m$^2$/s) it is speculated that some hydrolysis takes place in the fluid region of the L-DPPC monolayer. \\

When hydrolysis was allowed to proceed over more than an hour, large areas with distinctly slower enzyme diffusion (i.e. localized spots over a large area) were observed within the fluid region of L-DPPC monolayers (Fig. \ref{Figure6}A and \ref{Figure6}B). Since these areas were always detected near to the gel domains, we can speculate that they form as the debris of products generated at the gel/fluid boundary diffuses into the fluid region and accumulates. Although the existence of these so-called product domains has been previously reported, their structure and exact composition are largely unknown, but it seems clear that they contain, at least, calcium palmitate salts \cite{Maloney1993, Maloney1995, Grandbois2000}. \\
\begin{figure}[htb!]
    \begin{center}
	\includegraphics[width=8.5cm]{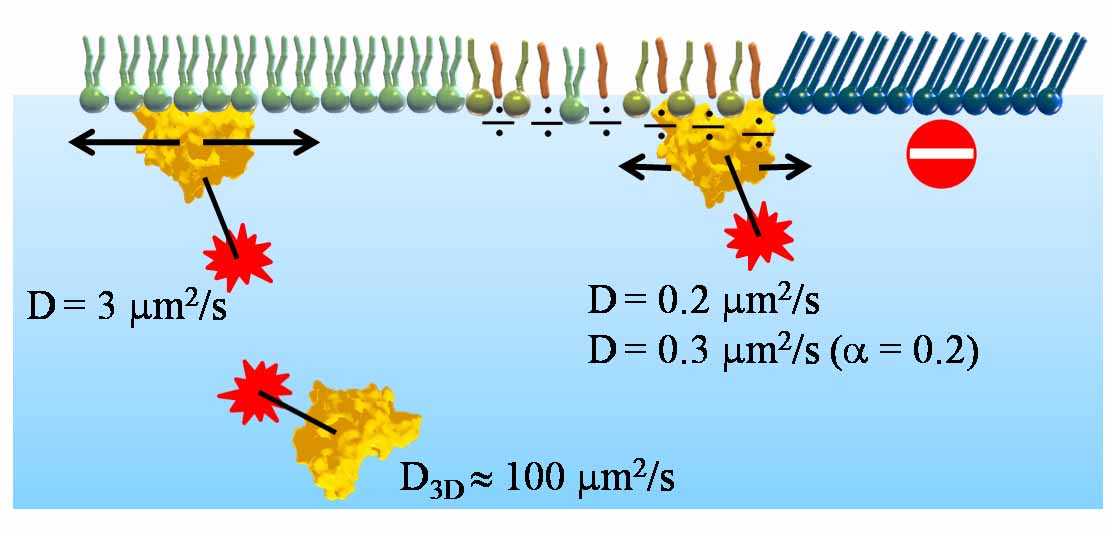}
	\parbox[c]{8cm}{ \caption{\textit{Cartoon representation showing typical modes of diffusion in different environments. For diffusion on the fluid region (lipids in green) the diffusion was generally normal with a diffusion coefficient of 3 $\mu$m$^2$/s. Enzymes located near L-DPPC gel domains showed more complex diffusion with e.g. transient trapping of the enzyme. After extended hydrolysis, areas enriched in hydrolysis product (molecules in olive green and red) showed enzyme diffusion which was significantly slower than on the fluid region. See text for details.}
	\label{Figure7}}}
    \end{center}
\end{figure}
On such product domains the enzyme molecules (2095 enzymes analyzed) presented diffusion with two characteristic diffusion coefficients, being D = 0.17 $\mu$m$^2$/s and D = 0.031 $\mu$m$^2$/s. Compared to the fluid L-DPPC region, the mobility of PLA$_2$-PDI is reduced by a factor of 20. Approximately half of the trajectories showed only the fast component, 1\% only the slow, while 48\% showed both periods of slow and fast diffusion. The M-LM image (Fig. \ref{FigureS3}) shows that the diffusion coefficients of PLA$_2$-PDI did not vary systematically within the product domain. Nevertheless, the H-LM shows a preferential localization of the enzyme for the product domain-gel domain interface (Fig. \ref{Figure6}C).\\

 In addition to the diffusion coefficients, the residence time (i.e. average time-length of the trajectories) for PLA$_2$-PDI was determined for the different systems. In all cases, a satisfactory fit was obtained by fitting the obtained distribution of trajectory duration to a two-component exponential decay. All systems contained the same relatively short residence time of 28 ms ($\pm$2.5 ms), as well as a more long-lived component which varied between 95-220 ms. The short residence time component is attributed to enzymes at the surface which did not bind specifically to the monolayer; e.g. enzyme at the interface with the binding motif (i-face) facing away from the monolayer. When within the product domain, the residence time of PLA$_2$-PDI molecules doubled. This further strengthens the hypothesis of a region enriched in negatively charged hydrolysis products where the enzyme binds more strongly. All diffusion coefficients and residence times are summarized in Table \ref{table1}. 


\begin{table*}
  \centering 
  \begin{tabular}{|c|c|c|c|c|c|c|}
\hline
 &\multicolumn{3}{|c|}{D ($\mu$m$^2$/s)} & \multicolumn{2}{|c|}{residence time (ms)} &number of\\
\cline{2-6}
monolayer system&\multicolumn{2}{|c|}{normal diffusion}&anomalous&short&short& trajectories\\
\cline{2-3}
&fast&slow&diffusion&component&component&analyzed\\
&component&component&&&&\\
\hline
L-DPPC&3.0&$<$0.26&-&28&95&1216\\
fluid region&&&&&&\\
\hline
L-DPPC near&0.27&0.025&0.27&31&220&515\\
domain&&&$\alpha=0.24$&&&\\
\hline
L-DPPC&0.17&0.031&-&27&190&2095\\
product domain&&&&&&\\
\hline
D-DPPC&4.6&$<0.16$&-&26&132&2971\\
fluid region&&&&&&\\
\hline
\end{tabular}
  \caption{Summary of diffusion coefficients and residence times of PLA$_2$-PDI}\label{table1}
\end{table*}
\section*{Conclusion}
The primary scope of this study was the direct visualization of activity and diffusion behavior of PLA$_2$ in a heterogeneous lipid environment. This was done by performing high resolution and high sensitivity time-resolved fluorescence imaging and single particle tracking of PLA$_2$-IB on L-DPPC monolayers during active hydrolysis. By tracking individual enzymes in the different phases of the substrate, differences in their mobility were related to differences in activity towards the different phases. All the experiments reported in the literature so far point to the fact that hydrolysis predominantly takes place at the domain boundaries \cite{Honger1996, OpDenKamp1974, Mouritsen2006}. Our results corroborate this notion trough direct visualization of enzyme activity. The results provide further evidence of a drastic change in the enzyme behavior at the gel-fluid boundary as hydrolysis progresses. The preferential binding of the enzyme to the gel-fluid boundary shown in the H-LM images on L-DPPC monolayer is not seen in absence of hydrolysis on a non-substrate D-DPPC monolayer. Moreover, the decreased mobility of PLA$_2$ in the fluid region of L-DPPC monolayers, relative to D-DPPC monolayers, indicates that some hydrolysis may also take place in the fluid region. \\

The advantage of SPT is clear in the context of this study as it allows us to distinguish between PLA$_2$Õs different modes of diffusion in different regions, and to determine the associated diffusion coefficients as well as residence times of the enzyme on the monolayer. In a condensed format, the enzyme was found to diffuse fast in the fluid regions of the L-DPPC monolayer, and slow near domain boundaries where hydrolysis predominantly takes place. The enzyme appears to have very low affinity on gel domains, presumably because the lipid packing is too dense for the enzyme to penetrate the domains (Fig. \ref{Figure7}).\\

\textbf{Acknowledgments:} We would like to thank Allan Svendsen (Novozymes A/S, Denmark) for supplying purified PLA$_2$. This project was supported by EU project BIOSCOPE and the KULeuven research fund (CREA2007). S.R. acknowledges the Portuguese Foundation for Science and Technology (FCT) for PhD grant SFRH/BD/27265/2006.

\small{

}

\end{document}